\RequirePackage[undo-recent-deprecations]{expl3}
\documentclass[a4paper,fleqn]{cas-dc}
\usepackage{soul}
\usepackage[sort&compress,numbers]{natbib}
\usepackage{lineno}
\usepackage{setspace}
 \usepackage[normalem]{ulem}

\def\tsc#1{\csdef{#1}{\textsc{\lowercase{#1}}\xspace}}
\tsc{WGM}
\tsc{QE}
\tsc{EP}
\tsc{PMS}
\tsc{BEC}
\tsc{DE}

\begin{document}
\let\WriteBookmarks\relax
\def\floatpagepagefraction{1}
\def\textpagefraction{.001}
\shorttitle{Mechanical Properties of Amorphous Carbon Nanotube and Nanoscroll}
\shortauthors{Pereira J\'unior \textit{et~al}.}

\title [mode = title]{A Reactive Molecular Dynamics Study on the Mechanical Properties of a Recently Synthesized Amorphous Carbon Monolayer Converted into a Nanotube/Nanoscroll}

\author[1]{Marcelo L. Pereira J\'unior}
\author[1]{ Wiliam F. Cunha}
\author[2,3]{ Douglas S. Galv\~ao}
\author[1]{ Luiz A. Ribeiro J\'unior}
\cormark[1]
\ead{ribeirojr@unb.br}

\address[1]{Institute of Physics, University of Bras\'ilia, Bras\'ilia, 70910-900, Brazil.}
\address[2]{Applied Physics Department, University of Campinas, Campinas, S\~ao Paulo, Brazil.}
\address[3]{Center for Computing in Engineering and Sciences, University of Campinas, Campinas, S\~ao Paulo, Brazil.}

\begin{abstract}
Recently, laser-assisted chemical vapor deposition was used to synthesize a free-standing, continuous, and stable monolayer amorphous carbon (MAC). MAC is a pure carbon structure composed of randomly distributed five, six, seven, and eight atom rings, which differs from disordered graphene. More recently, amorphous MAC-based nanotubes (a-CNT) and nanoscrolls (A-CNS) were proposed. In this work, we have investigated (through fully atomistic reactive molecular dynamics simulations) the mechanical properties and melting points of pristine and  a-CNT and a-CNS. Results showed that a-CNT and a-CNS have distinct elastic properties and fracture patterns concerning their pristine analogs. Both a-CNT and a-CNS presented a non-elastic regime before their total rupture, whereas the CNT and CNS undergo a direct conversion to fractured forms after a critical strain threshold. The critical strain for the fracture of the a-CNT and a-CNS are about 30\% and 25\%, respectively, and they are lower than the corresponding CNT and CNS cases. Although less resilient to tension, the amorphous tubular structures have similar thermal stability in relation to the pristine cases with melting points of 5500K, 6300K, 5100K, and 5900K for a-CNT, CNT, a-CNS, and CNS, respectively.  An interesting result is whereas the behavior of the pristine systems is substantially different depending on the system being a nanotube or a nanoscroll, thus indicating that the topology plays an important role, the same is not true for the amorphous version of the nanostructures, thus indicating that the structural disorder overcomes the topological features.
\end{abstract}



\begin{keywords}
Amorphous Carbon Nanotubes \sep Amorphous Carbon Nanoscrolls \sep Elastic Properties \sep Thermal Stability
\end{keywords}

\maketitle
\doublespacing

\section{Introduction}
Carbon-based nanomaterials have been used in numerous energy storage and conversion applications as good cost-benefit alternatives to conventional (silicon-based) technology \cite{bayatsarmadi2017recent,mauter2008environmental}. Many of their promises in revolutionizing these fields still need further refining for commercial-grade applications. Among such applications are worthy mention organic photovoltaics \cite{kippelen2009organic}, light-emitting diodes \cite{geffroy2006organic}, and thin-film transistors \cite{klauk2010organic}. The broad range of carbon-based nanomaterials employed in developing these applications includes graphene and its allotropes \cite{enyashin2011graphene}, carbon nanotubes \cite{iijima1991helical}, nanoscrolls \cite{viculis_S}, nanofibers \cite{hammel2004carbon}, and amorphous carbon materials \cite{robertson1986amorphous}. Particularly, carbon nanotubes (CNTs) are still subject of interest due to their remarkable electronic and structural properties, which have impacted several areas of nanotechnology \cite{haddon_ACR}. CNTs present distinct electronic properties, such as metallic and semiconducting behavior, depending on their symmetry \cite{popov2004carbon}. CNSs, in turn, are formed by the jelly or papyrus roll-like wrapping of a graphite sheet to form a tubular structure \cite{bacon_JAP,braga_NL,tomanek2002mesoscopic}. Recent synthesis efforts provided simple low-temperature routes to CNSs \cite{shioyama_C,viculis_S}.   

Amorphous carbons are families of carbon-based materials without long-range crystalline order \cite{robertson1986amorphous}. Several experimental efforts have been devoted to understanding the electronic and structural properties of amorphous CNTs (a-CNTs) for their applications in nanoelectronics \cite{ci2001crystallization,calderon2001hydrothermal,ci2003annealing,chen2012triple,lei2004large,hayashi2013reversible,bhowmick_RSC}. The synthesis and characterization of these materials were studied through simple fabrication procedures used for preparing highly ordered nanotube arrays \cite{lei2004large}. Electrical transport measurements on individual a-CNTs showed that their resistivity is of the same order as multiwalled CNTs despite their disordered microstructure \cite{hayashi2013reversible}. a-CNTs were also employed as potential candidates for water purification \cite{bhowmick_RSC}. It was reported that they can be used for adsorption and/or removal of hydroxyl group substituted aromatic compounds like Resorcinol as well as heavy metal ion like arsenic from water. 

From the theoretical point o view, the electronic properties of a-CNTs were studied by using a tight-binding model Hamiltonian \cite{rakitin_PRB}. Through this approach, it was obtained that strong electron-lattice coupling of states near the Fermi energy leads to the formation of a bandgap \cite{rakitin_PRB}. This model has also predicted that the bandgap in a-CNTs is stronger diameter dependent (inversely proportional) than for CNTs. In relation to CNS, some theoretical and experimental studies were performed to address possible synthetic routes \cite{shioyama_C,viculis_S,braga_NL,perim_FiM,xie2009controlled,perim_JAP,dravid1993buckytubes,dravid1993buckytubes,shioyama2003new}. There are only are very few studies in the literature on amorphous CNS (a-CNSs) \cite{gong2017graphene,yadav2012carbon}. To the best of our knowledge, up to now, there are no studies on the thermomechanical properties of a-CNTs and a-CNSs.

\begin{figure*}[pos=t!]
	\centering
	\includegraphics[width=0.7\linewidth]{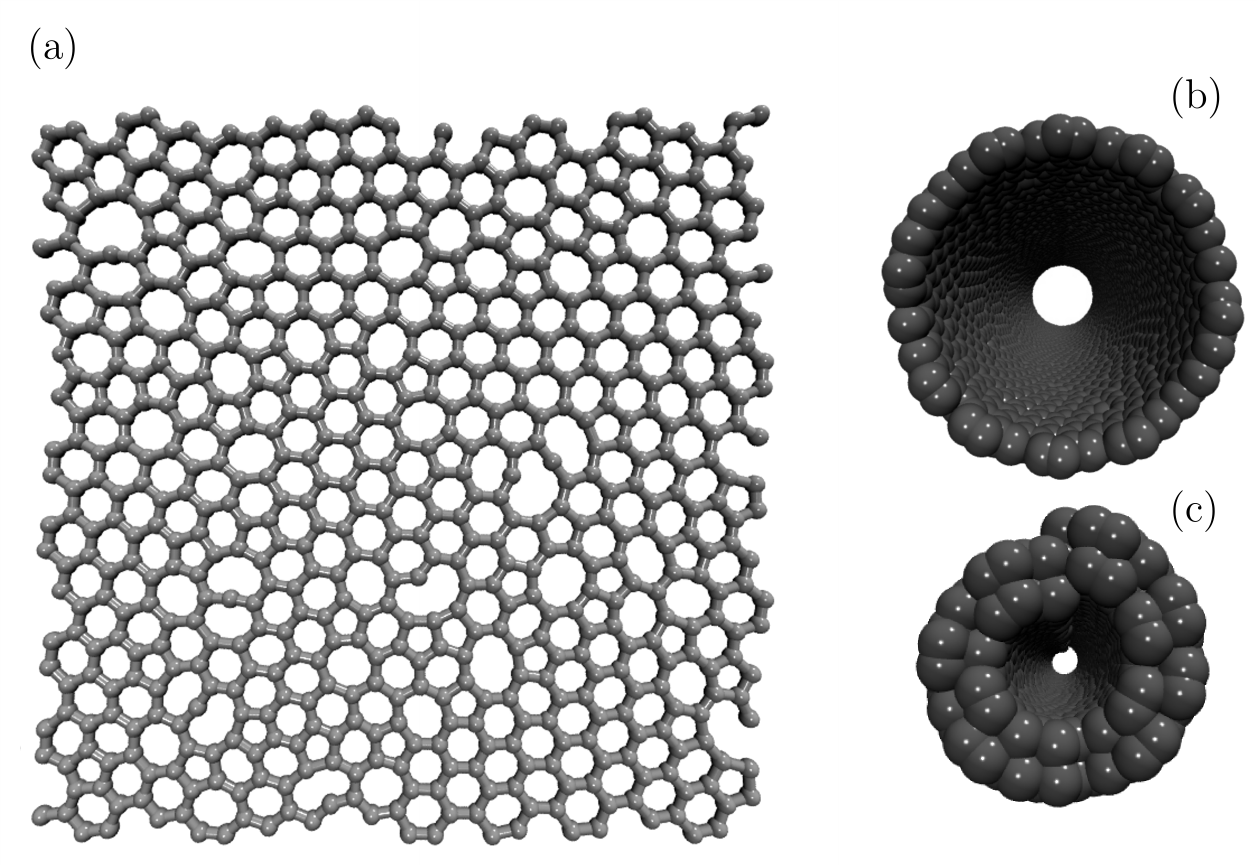}
	\caption{Structural models of the (a) MAC sheet and its related (b) a-CNT and (c) a-CNS atomistic structures.}
	\label{fig:systems}
\end{figure*}

Very recently, laser-assisted chemical vapor deposition was used to synthesize a free-standing, continuous, and stable monolayer amorphous carbon (MAC) \cite{toh_N}. MAC is a pure carbon structure composed of randomly distributed five, six, seven, and eight atom rings (see Figure \ref{fig:systems}(a)). Its lattice arrangement differs from disordered graphene. The findings have revealed that MAC has a high yield strength value with no crack propagation. Since the synthesis of MAC was achieved, its conversion to nanotubes and nanoscrolls emerges as a possibility.  

Motivated by the recent MAC synthesis \cite{toh_N} and by the proposition about the existence of tubes and scrolls formed from MAC membranes \cite{tromer2020optoelectronic} here the mechanical and thermal stability of MAC-based a-CNT and a-CNS were investigated in the framework of fully-atomistic reactive molecular dynamics simulations. The elastic properties of these model tubular materials were studied using the stress-strain relationship, and their thermal stability was analyzed by employing a heating ramp protocol. The results obtained here for these amorphous structures were contrasted against their related pristine ones.

\section{Details of Modeling}

We have performed fully-atomistic molecular dynamics (MD) simulations using the adaptive-interatomic reactive bond-order (AIREBO) \cite{stuart_JCP} potential as implemented in LAMMPS \cite{plimpton_JCP}. The main focus of the present work lies in providing a comparative study of the mechanical properties of pristine and MAC-based amorphous CNTs and CNSs. All these structures were generated by the Sculptor software suite \cite{humphrey_JMG}. The a-CNT was obtained by binding the two lateral edges of the MAC sheet \cite{toh_N,felix_JPCC}, whereas the a-CNS was formed by a spiral-like roll-up of the MAC sheet with open edges at both ends, with a rolling angle of 4$\pi$. Figure \ref{fig:systems} shows a representation of the MAC sheet (Figure \ref{fig:systems}(a)), its derived nanotube (Figure \ref{fig:systems}(b)), and nanoscroll (Figure \ref{fig:systems}(c)). The MAC atomistic structure used here to build our model a-CNT contains 610 atoms \cite{toh_N,felix_JPCC}. The yielded a-CNT has 25 \r{A} of diameter, 120 \r{A} of length, and it was formed by merging six MAC sheets (2 in the x-direction and 3 in the y-direction), totaling 3660 atoms. Here, a MAC sheet was considered as a unit cell. The a-CNS has also 120 \r{A} of length, but it possess 10 \r{A} and 16 \r{A} of internal and external diameters, respectively. The quantity of MAC sheets and the total number of atoms used to build the a-CNS are equal to the a-CNT case.

\begin{figure*}[pos=b]
	\centering
	\includegraphics[width=0.7\linewidth]{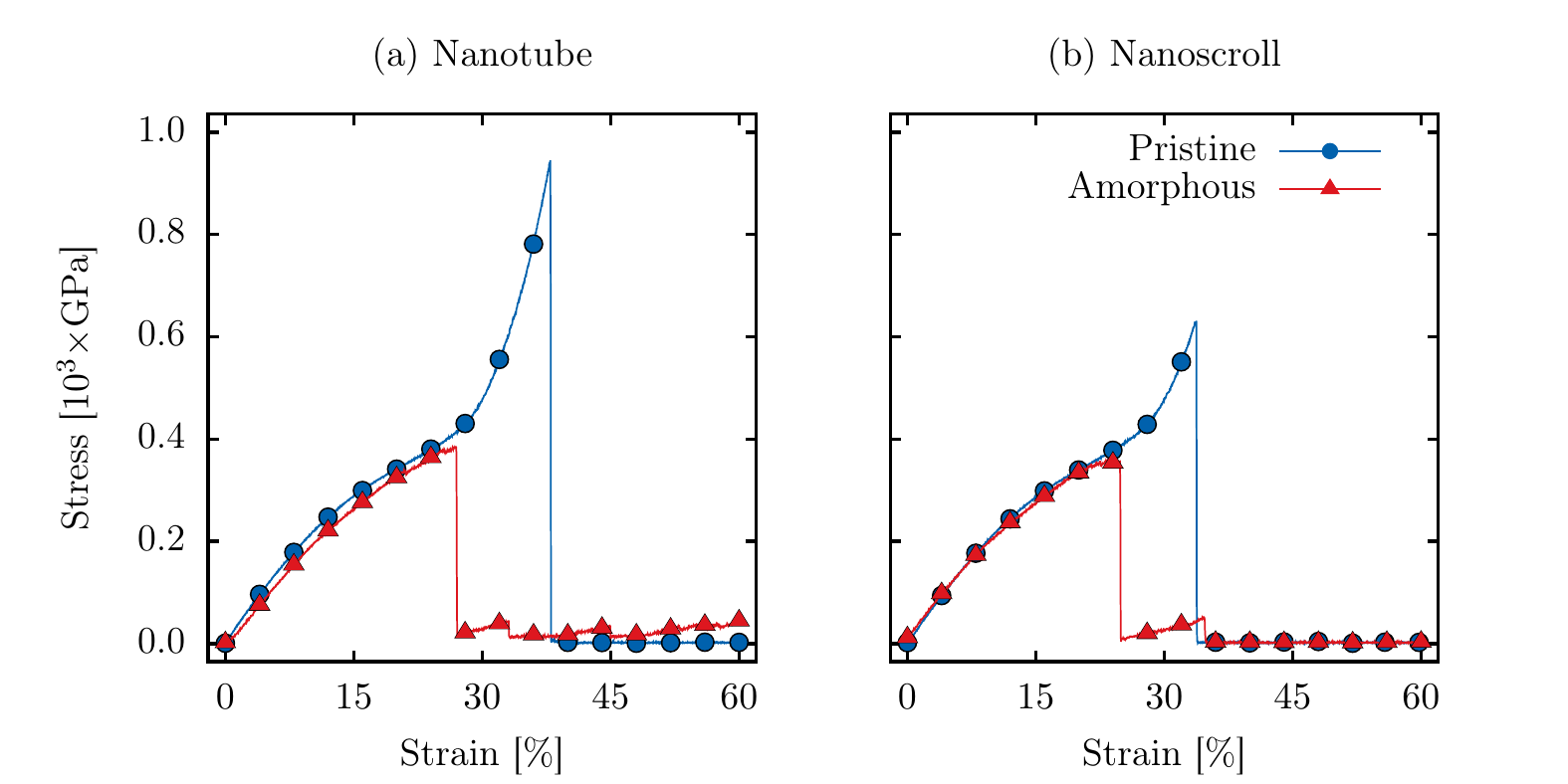}
	\caption{Stress-strain curves for (a) CNT (blue) and a-CNT (red) and (b) for CNS (blue) and a-CNS (red) at 300K, considering an uniaxial strain applied to the longitudinal direction. The structures were stretched at a constant elongation rate until total rupture.}
	\label{fig:ss}
\end{figure*}

Our computational approach is based on submitting the systems to tensile deformation to address their elastic properties as well as to a heating ramp process to obtain their estimated melting points. These analyses were performed by considering the virial stress tensor \cite{subramaniyan_IJSS}. We also calculated the von Mises stress \cite{mises_NGWZGM} as it is an important tool to describe the stretching dynamics and fracture patterns of the systems \cite{pereira_FC}.



As for the heating ramp procedure it consisted on a linear temperature increase in a NVT ensemble in which, a temperature range of from 0 up to 10000K was performed after a preliminary thermalization procedure within an NPT ensemble. The temperature ramp rate was set to approximately 200 K/ms. The MD snapshots and trajectories were obtained by using free visualization and analysis software VMD \cite{humphrey_JMG}. The equations of motion were numerically integrated using the velocity-Verlet integrator with a time-step of $0.1$ fs. We increased the tensile stress in the system by applying a uniaxial strain along the longitudinal direction of the tubular structures, for an engineering strain rate of $1.0 \times 10^{-7}$ fs$^{-1}$.

\section{Results}

Initially, we investigated the stress-strain behavior of both systems: the a-CNT (Figure \ref{fig:ss}(a)) and a-CNS (Figure \ref{fig:ss}(b)). In both figures, we present results of the structures derived from the pristine nanosheet in blue and of the amorphous one in red. In the two scenarios, the curves more or less overlap until a critical strain value is achieved. Such a critical strain value indicates a phase transition in the stress-strain dependence, which consists of the point of fracture of each structure. The fracture strain, for each case, is as follows: 27\%, 48\%, 24\%, and 34\% for a-CNT, CNT, a-CNS, and CNS, respectively.

\begin{table*}[pos=bh]
\begin{tabular}{|l|c|c|c|c|}
\hline
\multicolumn{1}{|c|}{System} & E {[}GPa{]} & D {[}GPa{]} & TS {[}GPa{]} & $\epsilon_F$ {[}\%{]} \\ \hline
MAC Nanotube                 & 2268.67     & 3001.13     & 384.97       & 26.72                 \\ \hline
Graphene Nanotube            & 2495.51     & 3927.15     & 944.36       & 37.96                 \\ \hline
MAC Nanoscroll               & 2476.61     & 3843.84    & 359.59       & 23.56                 \\ \hline
Graphene Nanoscroll          & 2493.82     & 4460.73     & 630.82       & 33.76                 \\ \hline
\end{tabular} \caption{Elastic properties for MAC and graphene nanotubes and nanoscrolls. E, D, TS, and $\epsilon_F$, refer to Young's Module, third-order elastic modulus, TS tensile strength, and critical strain, respectively.}
	\label{tab1}
\end{table*}

As expected, the general trend is that the pristine (defectless) structures can stand larger strain values without fracturing. Figure \ref{fig:ss}(a) shows that CNT is able to stand roughly 20\% more strain than its amorphous counterpart. Figure \ref{fig:ss}(b) compares the amorphous and pristine version of the nanoscroll. This figure shows that the pristine nanoscroll upholds 15\% more strain than its amorphous version. This trend is consistent with the fact that highly ordered carbon systems, such as graphene, will be tougher than its defective structures \cite{shekhawat2016toughness}. Such strength arises from the strong overlap of sp$^2$ hybrid orbitals. When the sites are uniformly close to each other, as is the case of the pristine scenario, the collective effect is that all the regions of the nanostructure exhibit the same strength. In the amorphous case, as two carbon cores approach one another, the relative strength between these two becomes stronger. However, as a consequence, the distance between a third core becomes larger and, therefore, the interaction is weaker. As the structure as a whole is only as strong as its weakest bond, it is expected that the amorphous tends to be weaker than the pristine systems. 

Another crucial feature of this set of simulations is observed through the comparison between Figures \ref{fig:ss}(a) and \ref{fig:ss}(b). We can observe, whereas the behavior of the pristine system is substantially different depending on the system being a nanotube or a nanoscroll, thus indicating that the topology plays an important role, the same is not true for the amorphous version of the nanostructures, thus indicating that the structural disorder overcomes the topological features (see also Table \ref{tab1}).

\begin{figure}
	\centering
	\includegraphics[width=0.8\linewidth]{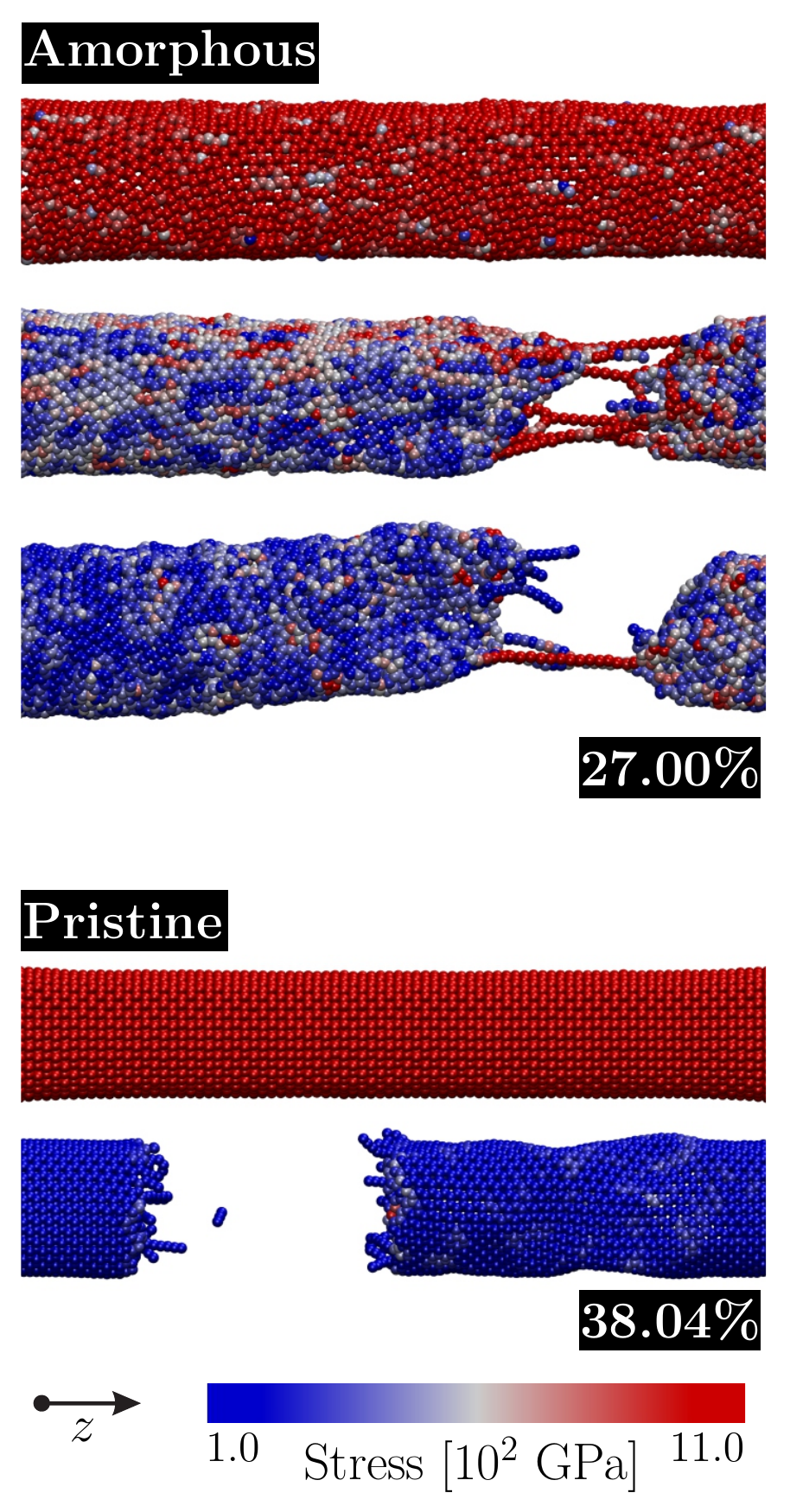}
	\caption{Representative MD snapshots for the fracture process for (top) a-CNT and (bottom) pristine CNT.}
	\label{fig:cnt}
\end{figure}

Table \ref{tab1} summarizes the set of elastic properties obtained from MD the simulations for the systems illustrated in Figure \ref{fig:systems}. It is worthwhile to stress that previous works have described the low-strain region as being quadratic due to the strong nonlinear behavior in the elastic regime \cite{lee_S,zandiatashbar_NC,zhong_JAP}. In this way, the stress-strain relationship is given by $\sigma = E\epsilon+D\epsilon^2$, where $\sigma$, $\epsilon$, $E$, and $D$ are the symmetric second Piola-Kirchhoff stress, uniaxial strain, Young's modulus, and third-order elastic modulus, respectively \cite{thurston_PR}. In the table, E stands for Young's Module, D is the third-order elastic modulus, TS the tensile strength and $\epsilon_F$ is the critical strain for the fracture. These values are numerical representations of the response of each material to the processes reported in Figure \ref{fig:ss}. As expected, the pristine structures presented higher Young's Modulus when contrasted with their amorphous analogs. However, these values are reasonably close. The more cohesive nature of CNTs confers to this material higher tensile strength and critical strain for the fracture. Due to the very nature of the system structures, the applied tensions are uniaxial and, therefore, Young's Modulus is a measure of the system stiffness. 

\begin{figure}
	\centering
	\includegraphics[width=0.8\linewidth]{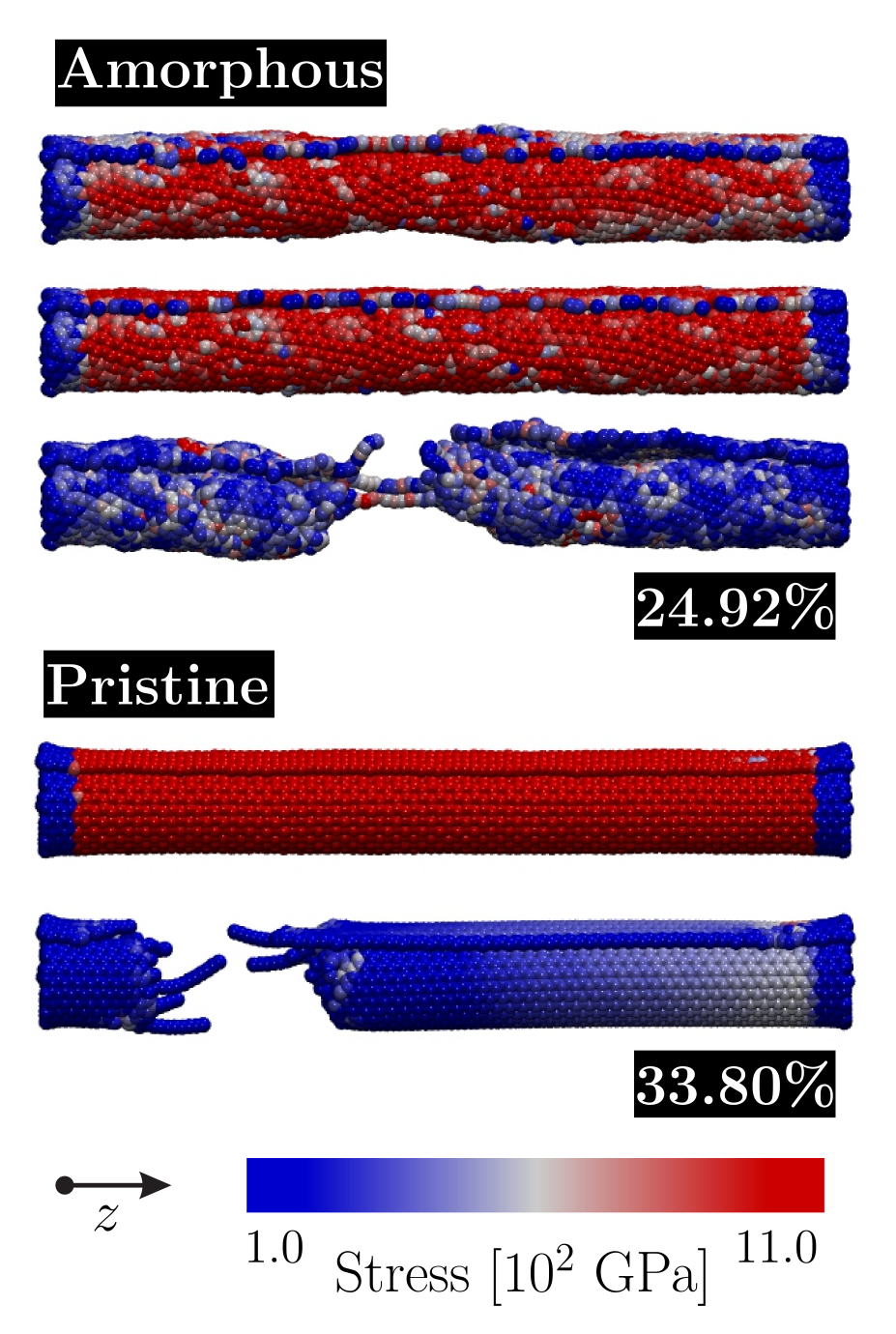}
	\caption{Representative MD snapshots for the fracture process for (top) a-CNS and (bottom) pristine CNS.}
	\label{fig:cns}
\end{figure}

In Figures \ref{fig:cnt} and \ref{fig:cns} we present representative MD snapshots of the fracturing processes of the different systems considered here and the von Mises stress distribution along the nanostructure. In these figures, the top and bottom sequence of panels illustrate the representative MD snapshots for the pristine and amorphous cases, respectively. Note that the a-CNT and a-CNS have a smaller fracture point in terms of strain in an approximate 1.4 ratio, which is consistent with the results presented in Figure \ref{fig:ss}. The trend of the amorphous structure to present smaller cutoff stress is preserved and rather similar among these two types of amorphous structures. Another clear trend to be observed is that before rupture, in both types of amorphous nanostructures, is the appearance of linear carbon chains (LACs) The same is not observed for pristine nanotubes and nanoscrolls. Due to the absence of porosity CNTs and CNS go from integrity to their fractured form abruptly. From these facts, one can again conclude that the structural disorder is of greater importance than the topological aspects.

\begin{figure*}
	\centering
	\includegraphics[width=0.7\linewidth]{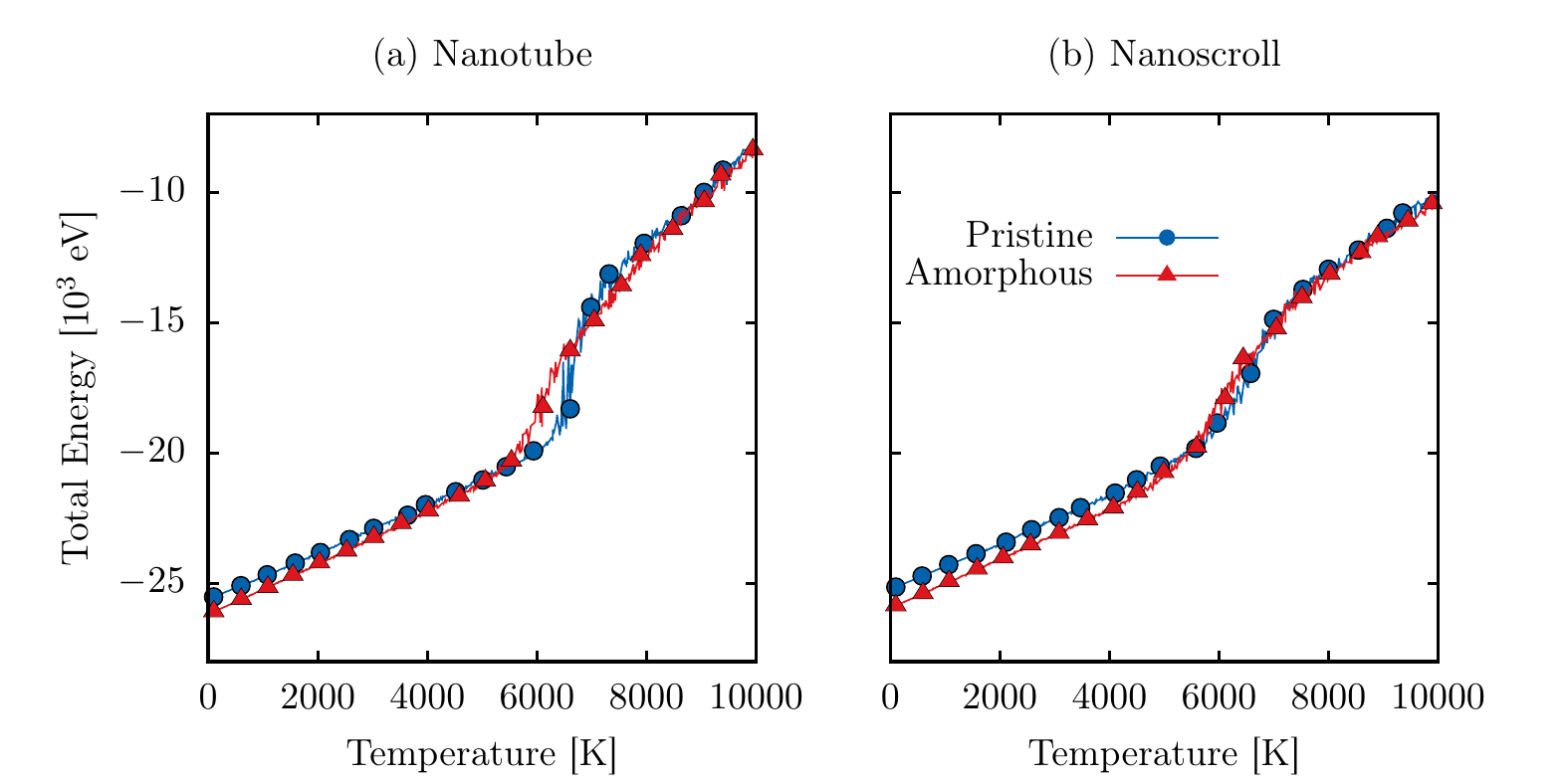}
	\caption{Total energy as a function of temperature for (a) nanotubes and (b) nanoscrolls. Amorphous structures are represented by red triangles and pristine ones by blue circles.}
	\label{fig:thermal}
\end{figure*}

\begin{figure}
	\centering
	\includegraphics[width=\linewidth]{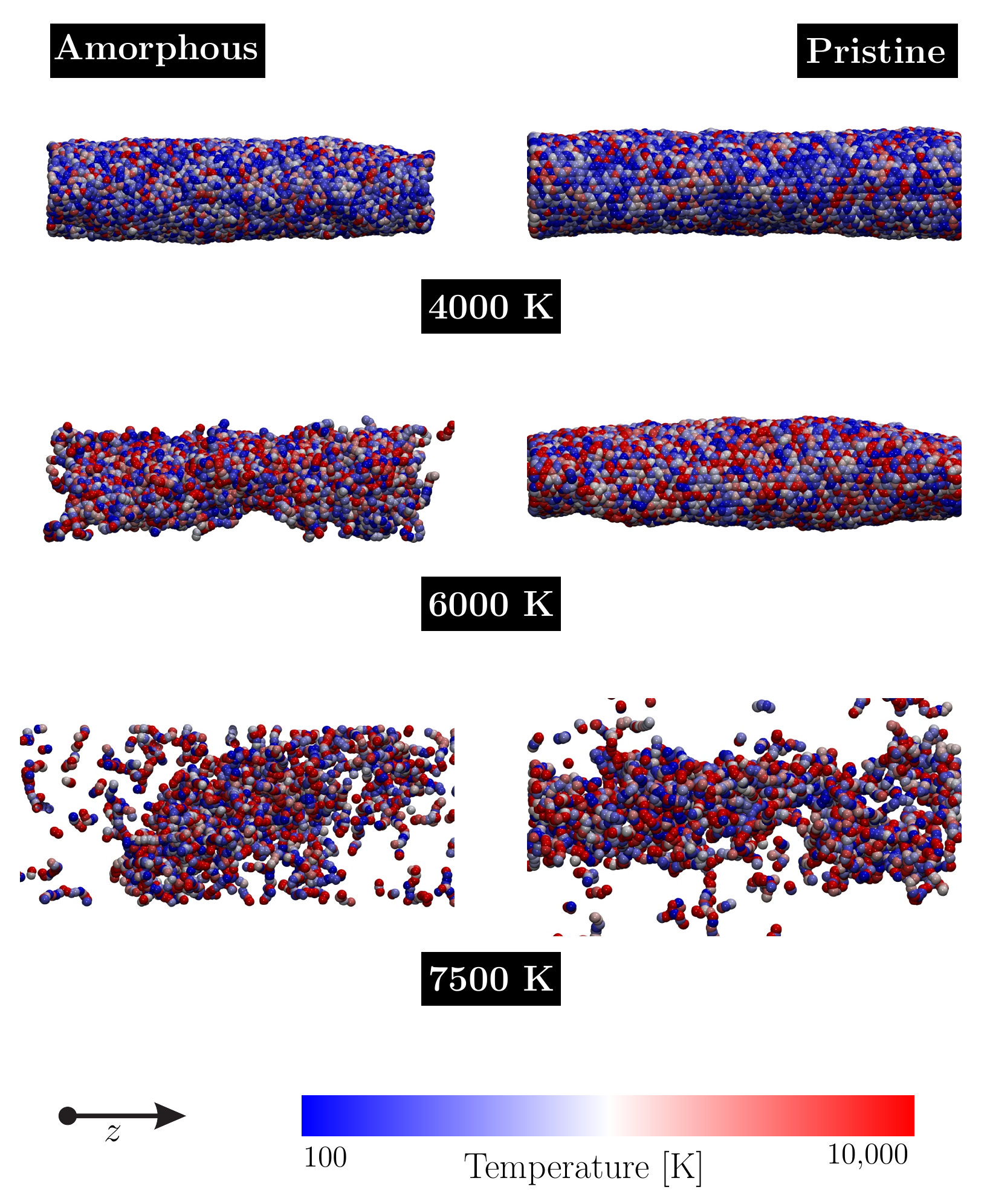}
	\caption{Representative MD snapshots of the heating simulations for (left) a-CNT and (right) pristine CNT.}
	\label{fig:heatingCNT}
\end{figure}

\begin{figure}
	\centering
	\includegraphics[width=\linewidth]{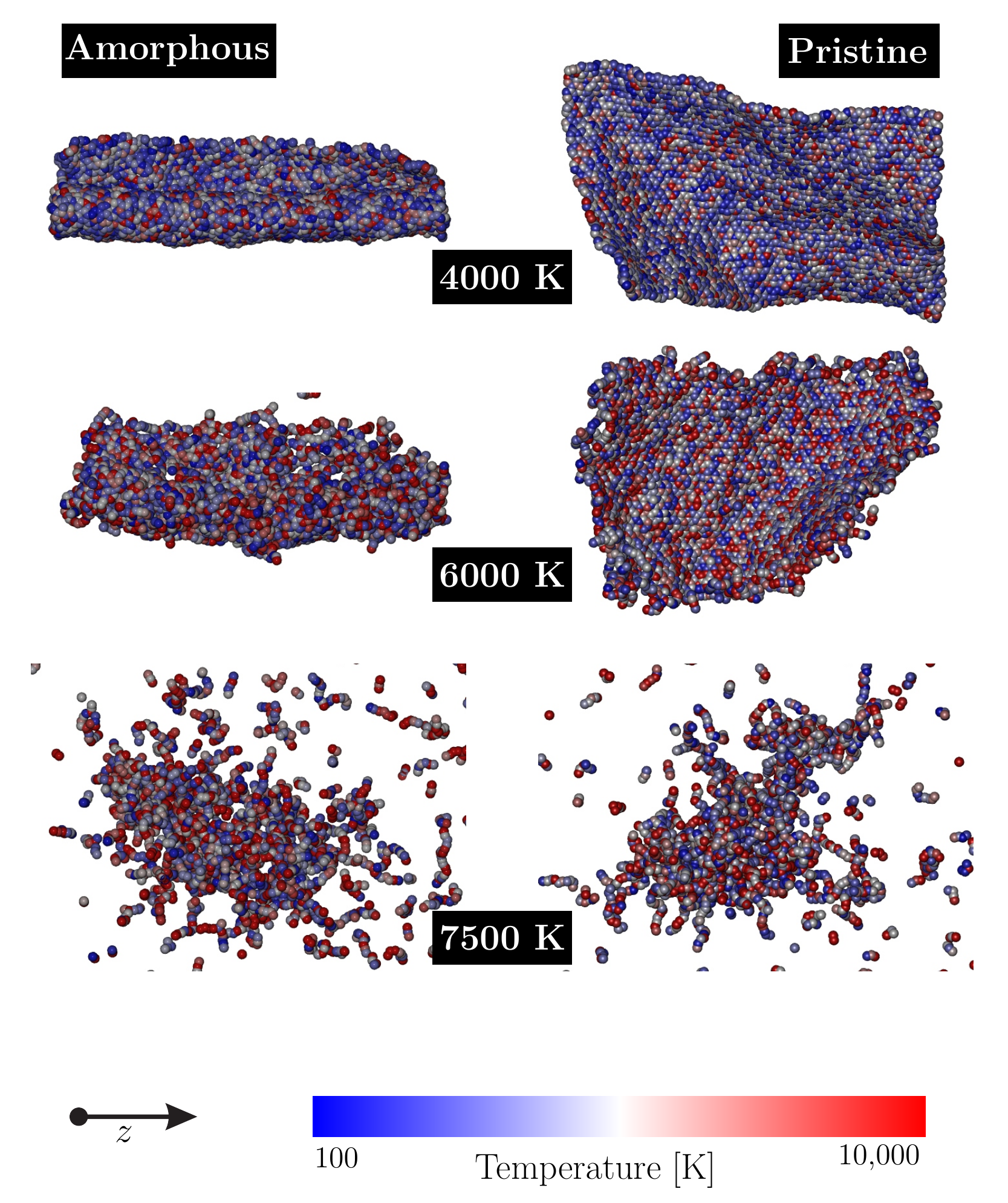}
	\caption{Representative MD snapshots of the heating simulations for (left) a-CNS and (right) pristine CNS.}
	\label{fig:heatingCNS}
\end{figure}

In Figure \ref{fig:thermal} we present the interplay between the total energy of the system and the thermal bath to it. The thermalization process for nanotubes (Figure \ref{fig:thermal}(a)) and nanoscrolls (Figure \ref{fig:thermal}(b)) is rather similar. For low temperatures, the curves are almost parallel, and the energy absorption of the amorphous cases are slightly smaller than the pristine ones. A feature that can be readily observed is that, as temperature increases, the relationship between these two quantities behavior for amorphous (red triangles) and pristine (blue circles) nanostructures is less important. This trend should, indeed, be expected, because higher temperature introduces disorder effects to the pristine chain, thus increasing the random degree of the atom movements, as is the case of the amorphous structure. A comparison between the two figures shows that the nanotubes absorb the thermal energy more efficiently than the nanoscrolls. This difference can be explained by the fact the scrolls have an open structure that allows converting the thermal energy into kinetic one. 

Figure \ref{fig:thermal} also allows us to have insights about the melting point of the nanostructures. As a general trend, the total energy increases linearly with the temperature and shows two well-defined regimes with different slopes. The discontinuity in these curves is suggestive of a phase transition from a solid to a gas-like phase. The transition between these two behaviors defines the melting point. This abrupt change in the slope of the curves is related to a gain in kinetic energy due to the higher atom velocities in the gas-like phase, which increases the total energy. Moreover, part of the harmonic and torsional energies in the solid phase was also converted into kinetic energy, which again contributes to the increase in the total energy of the system in the gas-like phase. The amorphous tubular structures have similar thermal stability when contrasted to the pristine cases with melting points of 5500K, 6300K, 5100K, and 5900K for a-CNT, CNT, a-CNS, and CNS, respectively.  

In Figures \ref{fig:heatingCNT} and \ref{fig:heatingCNS} we show representative MD snapshots for the heating simulations, for the nanotubes and nanoscrolls, respectively. In these figures, the left and the right sequence of panels depict the MD snapshots for the amorphous and pristine cases, respectively. In Figure \ref{fig:heatingCNT}, we can see that the a-CNT starts to lose its integrity, whereas the pristine is still stable at 6000K. At 7500K, both structures have reached their melting point, and a partially gas-like phase is observed. In the case of nanoscrolls (Figure \ref{fig:heatingCNS}), the pristine structure loses its structural integrity at 4000K, whereas the amorphous one at 5100K. As in the case of CNTs, at 7500K just a partially gas-like phase is observed, although more pronounced. From this set of simulations, it can be concluded that nanotubes are more thermally stable than nanoscrolls. As a general trend, one can also conclude that pristine structures are more stable than their amorphous counterparts, for the same reasons as the mechanical behavior discussed above. 

\section{Conclusions}

In summary, we have investigated (through fully atomistic reactive molecular dynamics simulations) the mechanical properties and melting point of pristine and amorphous carbon nanotubes and carbon nanoscrolls. The amorphous structures are based on the recently synthesized \cite{toh_N} stable monolayer amorphous carbon (MAC). Our results show that amorphous carbon nanotubes present lower tensile-strength than their corresponding pristine ones, which is expected since in general defectless structures are tougher than defective ones. The pristine structures exhibit abrupt fracture dynamics while the amorphous ones present intermediate states before complete failure. However, all structures present the appearance of linear atomic carbon chains before their rupture. The stress-strain curves for pristine/amorphous structures more or less overlap until the critical strain value is reached. The fracture strain, for each case, is as follows: 27\%, 48\%, 24\%, and 34\% for a-CNT, CNT, a-CNS, and CNS, respectively. An interesting result is whereas the behavior of the pristine systems is substantially different depending on the system being a nanotube or a nanoscroll, thus indicating that the topology plays an important role, the same is not true for the amorphous version of the nanostructures, thus indicating that the structural disorder outperform the topological features. The nanotubes of both symmetries present higher melting points than nanoscrolls, suggesting that the tubes are thermally more stable than the scrolls. In the case of the amorphous nanotube, a thermal critical value of 6000 K was obtained, at which point the pristine nanoscroll, for instance, had already lost its structural integrity. 

\appendix

\printcredits

\bibliographystyle{unsrt}
\bibliography{cas-refs}

\begin{thebibliography}{10}

\bibitem{bayatsarmadi2017recent}
Bita Bayatsarmadi, Yao Zheng, Anthony Vasileff, and Shi-Zhang Qiao.
\newblock Recent advances in atomic metal doping of carbon-based nanomaterials
  for energy conversion.
\newblock {\em Small}, 13(21):1700191, 2017.

\bibitem{mauter2008environmental}
Meagan~S Mauter and Menachem Elimelech.
\newblock Environmental applications of carbon-based nanomaterials.
\newblock {\em Environmental Science \& Technology}, 42(16):5843--5859, 2008.

\bibitem{kippelen2009organic}
Bernard Kippelen and Jean-Luc Br{\'e}das.
\newblock Organic photovoltaics.
\newblock {\em Energy \& Environmental Science}, 2(3):251--261, 2009.

\bibitem{geffroy2006organic}
Bernard Geffroy, Philippe Le~Roy, and Christophe Prat.
\newblock Organic light-emitting diode (oled) technology: materials, devices
  and display technologies.
\newblock {\em Polymer international}, 55(6):572--582, 2006.

\bibitem{klauk2010organic}
Hagen Klauk.
\newblock Organic thin-film transistors.
\newblock {\em Chemical Society Reviews}, 39(7):2643--2666, 2010.

\bibitem{enyashin2011graphene}
Andrey~N Enyashin and Alexander~L Ivanovskii.
\newblock Graphene allotropes.
\newblock {\em physica status solidi (b)}, 248(8):1879--1883, 2011.

\bibitem{iijima1991helical}
Sumio Iijima.
\newblock Helical microtubules of graphitic carbon.
\newblock {\em nature}, 354(6348):56--58, 1991.

\bibitem{viculis_S}
Lisa~M Viculis, Julia~J Mack, and Richard~B Kaner.
\newblock A chemical route to carbon nanoscrolls.(brevia).
\newblock {\em Science}, 299(5611):1361--1362, 2003.

\bibitem{hammel2004carbon}
E~Hammel, X~Tang, M~Trampert, T~Schmitt, K~Mauthner, A~Eder, and
  P~P{\"o}tschke.
\newblock Carbon nanofibers for composite applications.
\newblock {\em Carbon}, 42(5-6):1153--1158, 2004.

\bibitem{robertson1986amorphous}
John Robertson.
\newblock Amorphous carbon.
\newblock {\em Advances in Physics}, 35(4):317--374, 1986.

\bibitem{haddon_ACR}
Robert~C. Haddon.
\newblock Carbon nanotubes.
\newblock {\em Accounts of Chemical Research}, 35(12):997--997, 2002.
\newblock PMID: 12484786.

\bibitem{popov2004carbon}
Valentin~N Popov.
\newblock Carbon nanotubes: properties and application.
\newblock {\em Materials Science and Engineering: R: Reports}, 43(3):61--102,
  2004.

\bibitem{bacon_JAP}
Roger Bacon.
\newblock Growth, structure, and properties of graphite whiskers.
\newblock {\em Journal of Applied Physics}, 31(2):283--290, 1960.

\bibitem{braga_NL}
Scheila~F Braga, Vitor~R Coluci, Sergio~B Legoas, Ronaldo Giro, Douglas~S
  Galv{\~a}o, and Ray~H Baughman.
\newblock Structure and dynamics of carbon nanoscrolls.
\newblock {\em Nano letters}, 4(5):881--884, 2004.

\bibitem{tomanek2002mesoscopic}
David Tom{\'a}nek.
\newblock Mesoscopic origami with graphite: scrolls, nanotubes, peapods.
\newblock {\em Physica B: Condensed Matter}, 323(1-4):86--89, 2002.

\bibitem{shioyama_C}
H~Shioyama and T~Akita.
\newblock A new route to carbon nanotubes.
\newblock {\em Carbon (New York, NY)}, 41(1):179--181, 2003.

\bibitem{ci2001crystallization}
Lijie Ci, Bingqing Wei, Cailu Xu, Ji~Liang, Dehai Wu, Sishen Xie, Weiya Zhou,
  Yubao Li, Zuqin Liu, and Dongsheng Tang.
\newblock Crystallization behavior of the amorphous carbon nanotubes prepared
  by the cvd method.
\newblock {\em Journal of Crystal Growth}, 233(4):823--828, 2001.

\bibitem{calderon2001hydrothermal}
Jose~Maria Calderon~Moreno and Masahiro Yoshimura.
\newblock Hydrothermal processing of high-quality multiwall nanotubes from
  amorphous carbon.
\newblock {\em Journal of the American Chemical Society}, 123(4):741--742,
  2001.

\bibitem{ci2003annealing}
Lijie Ci, Hongwei Zhu, Bingqing Wei, Cailu Xu, and Dehai Wu.
\newblock Annealing amorphous carbon nanotubes for their application in
  hydrogen storage.
\newblock {\em Applied Surface Science}, 205(1-4):39--43, 2003.

\bibitem{chen2012triple}
Yuming Chen, Zhouguang Lu, Limin Zhou, Yiu-Wing Mai, and Haitao Huang.
\newblock Triple-coaxial electrospun amorphous carbon nanotubes with hollow
  graphitic carbon nanospheres for high-performance li ion batteries.
\newblock {\em Energy \& environmental science}, 5(7):7898--7902, 2012.

\bibitem{lei2004large}
Yong Lei, Kuan-Song Yeong, JTL Thong, and Wai-Kin Chim.
\newblock Large-scale ordered carbon nanotube arrays initiated from highly
  ordered catalyst arrays on silicon substrates.
\newblock {\em Chemistry of materials}, 16(14):2757--2761, 2004.

\bibitem{hayashi2013reversible}
Takuya Hayashi, Thomas~C O'Connor, Katsuhisa Higashiyama, Kohei Nishi, Kazunori
  Fujisawa, Hiroyuki Muramatsu, Yoong~Ahm Kim, Bobby~G Sumpter, Vincent
  Meunier, Mauricio Terrones, et~al.
\newblock A reversible strain-induced electrical conductivity in cup-stacked
  carbon nanotubes.
\newblock {\em Nanoscale}, 5(21):10212--10218, 2013.

\bibitem{bhowmick_RSC}
P~Bhowmick, D~Banerjee, S~Santra, D~Sen, B~Das, and KK~Chattopadhyay.
\newblock Amorphous carbon nanotubes as potent sorbents for removal of a
  phenolic derivative compound and arsenic: theoretical support of experimental
  findings.
\newblock {\em RSC advances}, 6(11):8913--8922, 2016.

\bibitem{rakitin_PRB}
A~Rakitin, C~Papadopoulos, and JM~Xu.
\newblock Electronic properties of amorphous carbon nanotubes.
\newblock {\em Physical Review B}, 61(8):5793, 2000.

\bibitem{perim_FiM}
Eric Perim, Leonardo~Dantas Machado, and Douglas~Soares Galvao.
\newblock A brief review on syntheses, structures, and applications of
  nanoscrolls.
\newblock {\em Frontiers in Materials}, 1:31, 2014.

\bibitem{xie2009controlled}
Xu~Xie, Long Ju, Xiaofeng Feng, Yinghui Sun, Ruifeng Zhou, Kai Liu, Shoushan
  Fan, Qunqing Li, and Kaili Jiang.
\newblock Controlled fabrication of high-quality carbon nanoscrolls from
  monolayer graphene.
\newblock {\em Nano letters}, 9(7):2565--2570, 2009.

\bibitem{perim_JAP}
Eric Perim, Ricardo Paupitz, and Douglas~S Galvao.
\newblock Controlled route to the fabrication of carbon and boron nitride
  nanoscrolls: A molecular dynamics investigation.
\newblock {\em Journal of Applied Physics}, 113(5):054306, 2013.

\bibitem{dravid1993buckytubes}
VP~Dravid, X~Lin, Y~Wang, XK~Wang, A~Yee, JB~Ketterson, and Robert~PH Chang.
\newblock Buckytubes and derivatives: their growth and implications for
  buckyball formation.
\newblock {\em Science}, 259(5101):1601--1604, 1993.

\bibitem{shioyama2003new}
H~Shioyama and T~Akita.
\newblock A new route to carbon nanotubes.
\newblock {\em Carbon (New York, NY)}, 41(1):179--181, 2003.

\bibitem{gong2017graphene}
Zhenbin Gong, Jing Shi, Bin Zhang, and Junyan Zhang.
\newblock Graphene nano scrolls responding to superlow friction of amorphous
  carbon.
\newblock {\em Carbon}, 116:310--317, 2017.

\bibitem{yadav2012carbon}
Prasad Yadav, Sambhaji Warule, Jyoti Jog, and Satishchandra Ogale.
\newblock Carbon nanoscrolls by pyrolysis of a polymer.
\newblock {\em Solid state communications}, 152(23):2092--2095, 2012.

\bibitem{toh_N}
Chee-Tat Toh, Hongji Zhang, Junhao Lin, Alexander~S Mayorov, Yun-Peng Wang,
  Carlo~M Orofeo, Darim~Badur Ferry, Henrik Andersen, Nurbek Kakenov, Zenglong
  Guo, et~al.
\newblock Synthesis and properties of free-standing monolayer amorphous carbon.
\newblock {\em Nature}, 577(7789):199--203, 2020.

\bibitem{tromer2020optoelectronic}
Raphael~M Tromer, Levi~C Felix, Luiz~A Ribeiro, and Douglas~S Galvao.
\newblock Optoelectronic properties of amorphous carbon-based nanotube and
  nanoscroll.
\newblock {\em arXiv preprint arXiv:2007.05062}, 2020.

\bibitem{stuart_JCP}
Steven~J Stuart, Alan~B Tutein, and Judith~A Harrison.
\newblock A reactive potential for hydrocarbons with intermolecular
  interactions.
\newblock {\em The Journal of chemical physics}, 112(14):6472--6486, 2000.

\bibitem{plimpton_JCP}
Steve Plimpton.
\newblock Fast parallel algorithms for short-range molecular dynamics.
\newblock {\em Journal of computational physics}, 117(1):1--19, 1995.

\bibitem{humphrey_JMG}
William Humphrey, Andrew Dalke, Klaus Schulten, et~al.
\newblock Vmd: visual molecular dynamics.
\newblock {\em Journal of molecular graphics}, 14(1):33--38, 1996.

\bibitem{felix_JPCC}
Levi~C Felix, Raphael~M Tromer, Pedro~AS Autreto, Luiz~A Ribeiro~Junior, and
  Douglas~S Galvao.
\newblock On the mechanical properties and thermal stability of a recently
  synthesized monolayer amorphous carbon.
\newblock {\em The Journal of Physical Chemistry C}, 124(27):14855--14860,
  2020.

\bibitem{subramaniyan_IJSS}
Arun~K Subramaniyan and CT~Sun.
\newblock Continuum interpretation of virial stress in molecular simulations.
\newblock {\em International Journal of Solids and Structures},
  45(14-15):4340--4346, 2008.

\bibitem{mises_NGWZGM}
R~v Mises.
\newblock Mechanik der festen k{\"o}rper im plastisch-deformablen zustand.
\newblock {\em Nachrichten von der Gesellschaft der Wissenschaften zu
  G{\"o}ttingen, Mathematisch-Physikalische Klasse}, 1913:582--592, 1913.

\bibitem{pereira_FC}
Marcelo Lopes~Pereira J{\'u}nior and Luiz Ant{\^o}nio~Ribeiro J{\'u}nior.
\newblock Thermomechanical insight into the stability of nanoporous graphene
  membranes.
\newblock {\em FlatChem}, 24:100196, 2020.

\bibitem{shekhawat2016toughness}
Ashivni Shekhawat and Robert~O Ritchie.
\newblock Toughness and strength of nanocrystalline graphene.
\newblock {\em Nature communications}, 7(1):1--8, 2016.

\bibitem{lee_S}
Changgu Lee, Xiaoding Wei, Jeffrey~W Kysar, and James Hone.
\newblock Measurement of the elastic properties and intrinsic strength of
  monolayer graphene.
\newblock {\em science}, 321(5887):385--388, 2008.

\bibitem{zandiatashbar_NC}
Ardavan Zandiatashbar, Gwan-Hyoung Lee, Sung~Joo An, Sunwoo Lee, Nithin Mathew,
  Mauricio Terrones, Takuya Hayashi, Catalin~R Picu, James Hone, and Nikhil
  Koratkar.
\newblock Effect of defects on the intrinsic strength and stiffness of
  graphene.
\newblock {\em Nature communications}, 5(1):1--9, 2014.

\bibitem{zhong_JAP}
Teng Zhong, Jiabin Li, and Kaiwang Zhang.
\newblock A molecular dynamics study of young’s modulus of multilayer
  graphene.
\newblock {\em Journal of Applied Physics}, 125(17):175110, 2019.

\bibitem{thurston_PR}
RN~Thurston and K~Brugger.
\newblock Third-order elastic constants and the velocity of small amplitude
  elastic waves in homogeneously stressed media.
\newblock {\em Physical Review}, 133(6A):A1604, 1964.

\end{thebibliography}

\end{document}